\documentclass[10pt,aps,pre,twocolumn,superscriptaddress,floatfix,showkeys,showpacs]{revtex4-2}
\usepackage{hyperref}
\usepackage{graphicx}
\usepackage{amsmath}
\usepackage[dvipsnames]{xcolor}
\hypersetup{colorlinks=true,breaklinks=true,allcolors=blue}

\usepackage{float}
\begin{document}

\title{ Feller diffusion in an interval: Inhomogeneous fluctuation–induced asymmetric escape}

\author{Prashanta Bauri}
\affiliation{
Department of Chemistry, Indian Institute of Technology Tirupati,
Yerpedu 517619, Andhra Pradesh, India
}

\author{Debasish Mondal}
\email{debasish@iittp.ac.in}
\affiliation{
Department of Chemistry, Indian Institute of Technology Tirupati,
Yerpedu 517619, Andhra Pradesh, India
}
\date{\today}

\begin{abstract}

We present an inhomogeneous fluctuation-induced asymmetric escape event of a Feller diffusion confined to a finite interval with two competing absorbing boundaries. The dynamics correspond to an overdamped Brownian motion in a shifted harmonic potential with state-dependent diffusivity. We set two exit points (equal potential energies), equidistant from the potential minima: an extinction site near the origin and an outbreak point. The multiplicative fluctuations are suppressed near the extinction boundary, biasing trajectories toward the outbreak state. The mean exit time exhibits a non-monotonic dependence on the initial condition and the drift-to-noise strength ratio, attaining a maximum when the particle is initialized toward the extinction (low-noise) site. The outbreak possibility is more likely even if the process started with a small initial bias towards the low-noise boundary. The spatial location of the lowest coefficient of variation (CV) is nicely corroborated by the maximum exit time, which is the hallmark of a stochastic escape from an interval. The fluctuation in escape time is found to dominate its mean as a non-trivial function of the drift-to-noise strength and the initial spatial bias. The observed asymmetries in escape also shape the speed–accuracy trade-off in stochastic decision-making events.

\end{abstract}


\maketitle

\section{Introduction}
\label{sec1}
 Stochastic movements with spatially varying diffusivity, in which the strength of the noise fluctuation depends explicitly on the system's current state (e.g., position, velocity, or energy), arise naturally in many biological and physical processes \cite{durr_1982JSP,horsthemke_1984book,lubensky_2007PRE,hummer_2008prl,gopinathan_2010,wehr_2016rpp,volpe_2016rmp,rieger_2019}. A simple example is position-dependent tRNA diffusion inside the ribosome, where the nonuniform interaction arises from the electrostatic interaction with ribosome components \cite{puglisi_2004NSMB,tung_2005pnas,scott_2007mc, Paul_2019}. Similarly, a particle in geometric confinement undergoes position-dependent diffusion, e.g., a colloidal particle in porous media \cite{rice_2000PRE,grier_2001EPL,ramirez_2011JCP} and a particle trapped in vesicles or between two parallel membranes \cite{ostrowsky_1996PRB,ostrowsky_2001EPL}. The friction coefficient of the particle, and hence its diffusion coefficient, may vary due to the hydrodynamic interaction with its distance from the wall \cite{ostrowsky_2001EPL,ostrowsky_2002physica}. Molecular transport in crowded environments, such as biological cytoplasm or porous media, often exhibits anomalous diffusion and is strongly influenced by spatial heterogeneity \cite{sokolov_2012sm,franosch_2013rpp}. In population dynamics, the growth (extinction or persistence) of the population under demographic or environmental variability depends on its current population size \cite{scott_1991em,erik_2003book,hastings_2008nature,lande_2022book}. Multiplicative noise frequently arises in space-dependent problems \cite{horsthemke_1984book, MARCHESONI_1984cpl, MAMunoz_1997prl, SKBanik_2008jcp, RYuan_2014jcp, RMetzler_2015jcp, DEMarakov_2017jcp, YChen_2019jcp, somrita_2020JCP}. Unlike additive noise (independent of the system's state), multiplicative noise can induce an effective potential landscape (effective drift) that may alter both stationary and non-stationary dynamics of the process. This leads to many fascinating fluctuation-driven phenomena, like noise-induced transitions \cite{lefever_1977pla,lefever_1978zfpbcm,inaba_1980ptp,toral_1994prl, dsray_2010jcp}, asymmetric localization \cite{pulak_2007pre,dsray_2011jcp}, and resonant activation \cite{bcbag_2014jcp, jonathan_1992prl}.\\

Fluctuation-induced escape from an interval with two competing absorbing boundaries is a control framework for understanding noise-driven stochastic decision-making processes \cite{julicher_2015prl}. Now, the presence of a spatially inhomogeneous diffusivity ($D(x)$) could introduce nontrivial asymmetries in the escape events. The likelihood, as well as the rate of escape to either boundary, is no longer influenced solely by the symmetry of the deterministic force field. In such a situation, a boundary with low diffusivity can act as a kinetically frozen bottleneck, reducing the escape rate and the likelihood of using that exit channel. On the other hand, the random movements of the particle (or process) near the escape point, with comparatively higher noise fluctuations, are quite fast. Such enhanced diffusivity increases the possibility of escape through that exit channel, essentially creating an asymmetric bias in the splitting probabilities. For example, consider a competitive extinction-outbreak event in population dynamics in which the underlying deterministic force field favors a symmetric outcome. Now, the presence of an inhomogeneous noise structure in the dynamics may alter the balance of escape probabilities and the escape rate in a nontrivial manner. Therefore, a thorough understanding of this asymmetric escape would be useful for predicting the outcome percentage and for designing desired control strategies for a non-equilibrium decision-making process with inherent spatial heterogeneity.

A Feller process is a special kind of one-dimensional Markov process with a linear drift (harmonic force) and a linear space-dependent fluctuation that vanishes at the origin \cite{Azaele_2006,feller_1951,feller_1952,feller_1954,masoliver_2012,masoliver_2014PRE,masoliver_2016,somrita_2022}.
The process is useful for modeling biophysical phenomena in which the realization of a negative value is forbidden, and the randomness depends on the current state. Capocelli and Ricciardi mathematically formulated the Feller process~\cite{capocelli_1973,capocelli_1974} with a one-dimensional space-dependent diffusion model. The process can be considered an alternative to the popular Lotka-Volterra model \cite{lotka_1925, Volterra_1928IJMS}, which describes the evolution of species in a fluctuating environment. The process has various applications in theoretical biology and neurology~\cite{wolf_2010, bibbona2010,d2018two} and population growth~\cite{azaele_2010, Azaele_2006,visco2010switching}, such as the spreading of a disease (medical) \cite{hilhorst_1999epjb}, maximal segmental score (genetics), and movement patterns of various animals (biology) \cite{reynolds2007}. A Feller process can even be used in economics and financial markets~\cite{heston1993, dragulescu2002probability, richmond2004langevin, sornette2003}. 
For example, in mathematical finance, the Cox-Ingersoll-Ross (CIR) model, which describes the time evolution of the interest rate in a volatile market, is often considered a Feller process \cite{cox1985aa}. In physics, the Feller process is often used \cite{woyczynski2001} to model Brownian motion with varying friction, phase transitions, and heat conduction in materials with nonuniform properties. \\

The long-term behavior of Feller diffusion and associated level-crossing phenomena, such as the first-passage time with a single absorbing boundary, has been studied recently by Masoliver and others \cite{masoliver_2012, satya_2013}. S. Ray, in 2022, showed the effect of stochastic resetting on its first-passage properties in the presence of a single escape window \cite{somrita_2022}. However, the stochastic escape of a Feller process from a finite interval with two competing exit channels remains unexplored. As mentioned earlier, escapes from a Feller process occur in the presence of spatially heterogeneous diffusivity. Thus, the interplay between the noise-induced drift and nonlinear diffusivity is expected to generate qualitatively new escape asymmetries. We address this issue by exploring escape rates, splitting probabilities, and the related speed-accuracy interplay. 

\section{The model and Stationary distribution}
\label{sec2}

 To begin with, we model a Feller diffusion that represents the dynamics of an overdamped Brownian motion confined to a shifted harmonic potential, with inhomogeneous diffusivity following the spirit of \cite{Azaele_2006,feller_1951, feller_1952, feller_1954, masoliver_2012,  masoliver_2012,masoliver_2014PRE,masoliver_2016, somrita_2022}.
A stochastic differential equation (SDE) of the process $Y(t')_{t'\in[0,\infty)}$ can be expressed as (It\^{o} interpretation):
\begin{equation}
   dY(t')=\alpha\left[\beta-Y(t')\right]dt'+\sigma\sqrt{D_0Y(t')}\,dW(t'),
    \label{eq1}
\end{equation}

where $\alpha$ (dimension $[T^{-1}]$), $\beta$ (dimension of the state variable, e.g., $[L]$), and $\sigma$ (dimension $[L^{1/2}T^{-1/2}]$) are positive constants, and $D_0$ is a dimensionless parameter that controls the fluctuation amplitude. Here $W(t')$ is a standard Wiener process with $\langle dW(t')\rangle=0$ and $\langle dW(t')dW(t'')\rangle=\delta(t'-t'')dt'dt''$.

Removing the dimensions simplifies the equation and reveals universal scaling relationships across diverse systems, irrespective of the actual values of the physical parameters. In this spirit, we scale the time and the spatial coordinate of the process by considering the dimensionless variables:
$t=\alpha t'$ and $x =\frac{2\alpha}{\sigma^2} Y$.
Applying these scaled variables, Eq.~\ref{eq1} can be rearranged as:
\begin{equation}
    \frac{d x(t)}{d t}= -\left[x(t)-\theta\right]+\sqrt{2D(x(t))}\,\xi(t),
    \label{eqn2}
\end{equation}
where $\theta=\frac{2\alpha\beta}{\sigma^2}$ and $D(x)=D_0x$. The rescaled Gaussian white noise $\xi(t)$ satisfies $\langle\xi(t)\rangle=0$ and $\langle\xi(t)\xi(t')\rangle=\delta(t-t')$. Note that Eq.~\ref{eqn2} is the dimensionless overdamped Langevin description of the process. Both $\theta$ and $D_0$ are considered positive quantities.
An alternative description of Eq.~\ref{eqn2}, in terms of the probability density function (PDF), can be presented by the following  Fokker-Planck Equation (FPE) \cite{risken_1989}:
\begin{equation}
    \frac{\partial p(x,t|x_0)}{\partial t}=-\frac{\partial}{\partial x}\Big[(\theta-x)p(x,t|x_0)\Big]+D_0\frac{\partial^2}{\partial x^2}\Big[xp(x,t|x_0)\Big],
\label{eqn3}
\end{equation}
where $p(x,t|x_0)$ is the conditional PDF of the process with the initial condition $p(x,0|x_0)=\delta(x-x_0)$; $x_0$ denotes the initial state.
\begin{figure}
    \centering
    \includegraphics[width=1.0\linewidth]{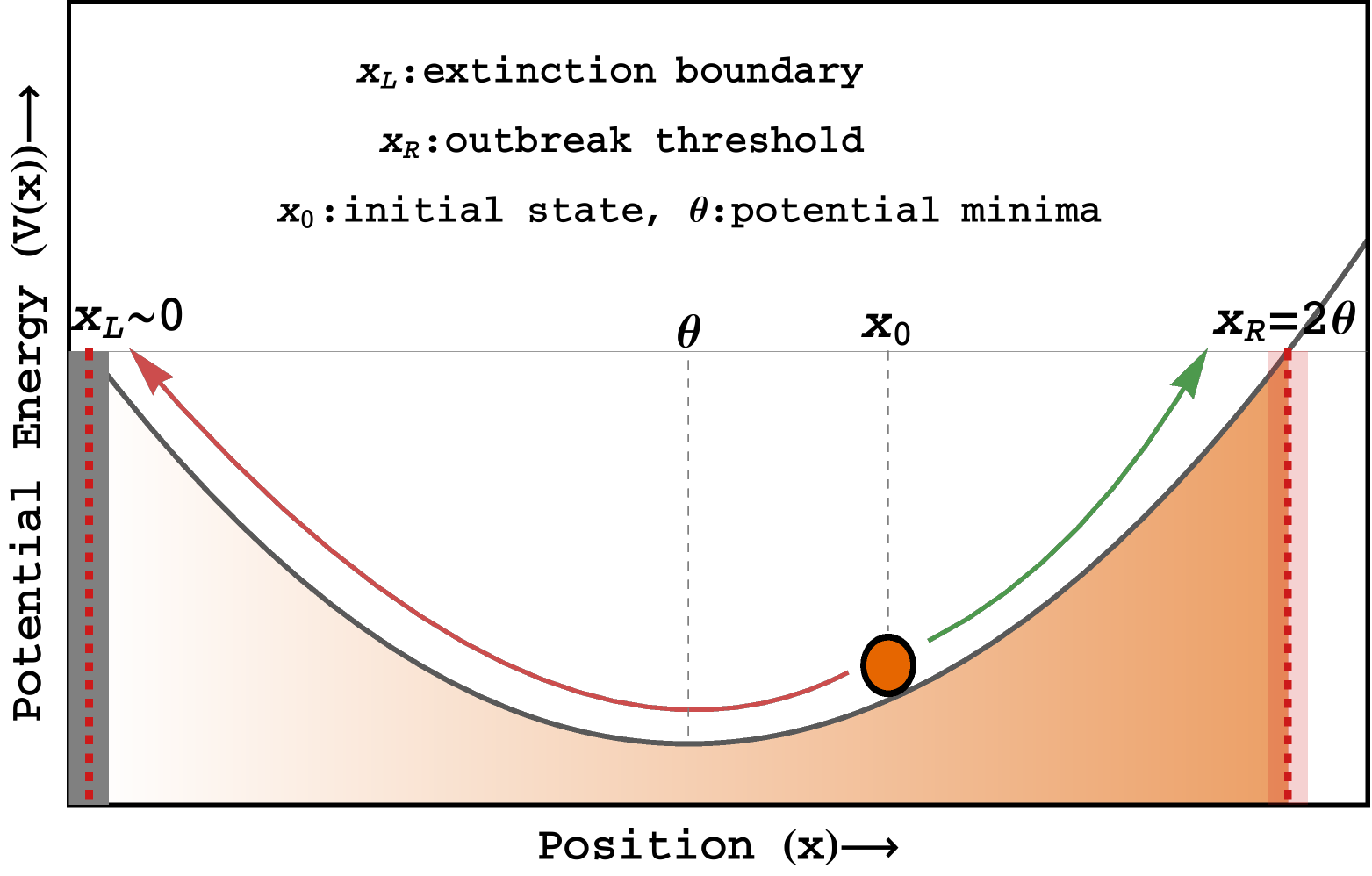}
    \caption{Schematic diagram of the potential energy set-up and location of the absorbing boundaries of the Feller process under study. Two absorbing boundaries are located at $x_L$ (left) and $x_R$ (right).}
    \label{fig1}
\end{figure}
Due to frozen diffusivity, the Fokker-Planck operator becomes degenerate at the origin, and thus Eq.~\ref{eqn3} exposes a singular/critical point at $x=0$. For \(x>0\), the operator is regular. Depending on the boundary classification at $x=0$ (controlled by $\theta/D_0$ for the It\^{o} dynamics considered here), trajectories started at $x_0>0$ remain on the positive side of the origin. Since the system exhibits position-dependent diffusivity, the fluctuation tends to zero as the particle approaches the origin and increases as it moves away from the origin. \\


 The transition probability density $p(x,t|x_0)$ corresponds to Eq.~\ref{eqn3} was obtained by Feller himself and others using different approaches \cite{feller_1951,masoliver_2012} and has a form:
\begin{align}
p(x,t|x_0)
 &= \frac{1}
     {D_0(1-e^{-t})}\left(\frac{x}{x_0 e^{-t}}\right)^{\frac{\frac{\theta}{D_0} - 1}{2}} \notag\\
 &\times \exp\left[-\frac{x+x_0 e^{-t}}{D_0(1-e^{-t})}\right] I_{\frac{\theta}{D_0} - 1}\left(\frac{2\sqrt{x x_0 e^{-t}}}{D_0(1-e^{-t})} \right),
\label{eqn4}
\end{align}
where $I_{\frac{\theta}{D_0}-1}(z)$ is a modified Bessel function of the first kind \cite{Arfken_1985,soni_1966}:
\begin{equation}
    I_{\frac{\theta}{D_0}-1}(z)=\sum_{n=0}^{\infty}\frac{1}{n!\Gamma(n+\frac{\theta}{D_0})}\left(\frac{z}{2}\right)^{\frac{\theta}{D_0}-1+2n}.
\end{equation}
One can thus find the stationary PDF either by considering a long time limit $t\rightarrow{\infty}$ in Eq.~\ref{eqn4} or by applying the stationary condition to the FPE (Eq.~\ref{eqn3}) as $\frac{\partial p_{st} }{\partial t}=0$. One can obtain the stationary probability distribution function ($p_{st}(x)$) as:
\begin{equation}
   p_{st}(x)=\mathcal{N}x^{\frac{\theta}{D_0}-1}e^{-\frac{x}{D_0}},
   \label{eqn6}
\end{equation}
where $\mathcal{N}$ is the normalization constant and has the form
 $\mathcal{N}^{-1}=D_0^{\frac{\theta}{D_0}}\Gamma\left(\frac{\theta}{D_0}\right)$. Also, $p_{st}(x)$ is only feasible when $\frac{\theta}{D_0}>0$.
\begin{figure}[!h]
\includegraphics[width=\linewidth]{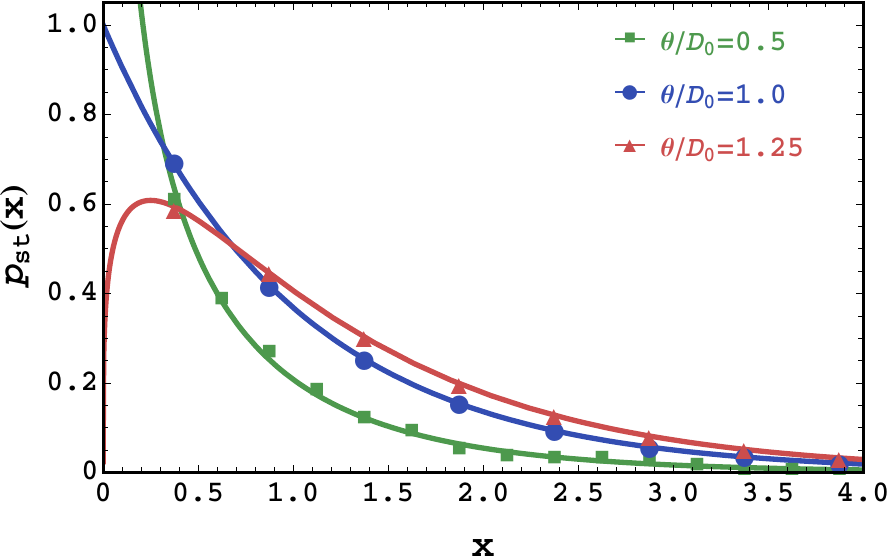}
 \caption{Variation of the stationary probability distribution function, $p_{st}(x)$ as a function of space coordinate $x$ for three different regimes of $\frac{\theta}{D_0}$ ratio. If $\frac{\theta}{D_0}\le1$, then the origin is accessible (green and blue lines). In the case of $\frac{\theta}{D_0}>1$, the origin is not accessible (red colored line). Lines are obtained analytically following Eq.~\ref{eqn6} and the points are obtained from the Langevin dynamics simulation of Eq.~\ref{eqn2}.}
\label{fig2}
\end{figure}
Eq.~\ref{eqn4} shows that as time progresses, the distribution converges to a gamma distribution \cite{Arfken_1985}. In Fig.~\ref{fig2}, we present a variation of $p_{st}(x)$ as a function of the spatial coordinate ($x$) for different $\frac{\theta}{D_0}$ ratios. As $\theta$ determines the depth of the deterministic potential well and $D_0$ indicates the strength of the stochastic force, we denote $\frac{\theta}{D_0}$ as the drift-to-noise ratio. The distribution shows a distinct noise-induced transition when the $\frac{\theta}{D_0}$ ratio is above or below unity. 
 In the range of $0 < \frac{\theta}{D_0}<1$, the distribution diverges at $x \sim 0$. For $\frac{\theta}{D_0} = 1$, the origin is accessible with a finite probability. For the It\^{o} Feller dynamics of Eq.~\ref{eqn2}, when $\frac{\theta}{D_0}>1$ the point $x=0$ becomes an entrance (inaccessible) boundary, i.e., trajectories started at $x_0>0$ do not reach the origin. Physically, the freezing of the diffusivity near the origin together with the drift towards the stable point $\theta$ suppresses approach to $x=0$ in this regime. We use this information to place absorbing boundaries in the latter part of this study.
\section{Extinction and Outbreak Boundaries, and Simulation details}
\label{sec3}

The inhomogeneous fluctuations lead to the observed non-equilibrium transition (Fig.~\ref{fig2}) at different drift-to-noise ratios. Therefore, the heterogeneous noise structure is expected to affect escape timescales from an interval in a nontrivial manner as well. Here, we consider two fixed absorbing ends at $x_L$ and $x_R$, as shown in Fig.~\ref{fig1}. Henceforth, we refer to $x_L$ as the extinction boundary and to $ x_R$ as the outbreak point, respectively. In principle, for a bounded Feller diffusion process, one is free to place absorbing boundaries in any position on the positive side of the spatial coordinate ($ x_{abs} >0$). We decide to place them at $x_L\sim 0$ ($~ 10^{-4}$) and $x_R=2\theta$. 
If the escape process is initiated from the potential minimum ($x_0=\theta$), then the choice of such boundary positions makes the deterministic part of Eq.~\ref{eqn2} spatially approximately symmetric. Now, for a similar homogeneous diffusion setup, one can expect equal splitting probabilities for the outbreak (here, doubling in $x_0$) and extinction events. However, because the Feller process has spatially inhomogeneous diffusivity, fluctuations are high when the particle is far from the origin and freeze near the origin. Therefore, such space-dependent noise is expected to yield non-trivial outcomes for similar variations in splitting probability and mean escape rate. \\

Whenever necessary, we numerically solve the Langevin dynamics (Eq.~\ref{eqn2}) using an improved Euler method \cite{xavier_1994} with a time step $\Delta t = 10^{-4}$, and with appropriate absorbing boundary conditions to find the stationary PDF and other physical observables to support theoretical predictions. We employ a Box-Muller algorithm to generate the Gaussian noise \cite{muller_1958}. We generate a large number of trajectories ($\sim 2-5\times10^5$ ) for statistical averaging. Unless mentioned otherwise, we set $D_0=1.0$, $x_L=10^{-4}$, and $x_R=2\theta$. The agreement between the numerical simulation data of the $p_{st}(x)$, $\langle\tau\rangle$, and splitting probabilities and the exact analytical estimations, as shown in Figs.~\ref{fig2}-\ref{fig4}, \ref{fig6}, and \ref{fig8} justifies the choice of the time step and other statistical averaging parameters.
\section{ Mean Escape time and Coefficient of Variation}
\label{sec4}
The mean escape time ($\langle \tau \rangle$), popularly known as the mean first passage time (MFPT), is the average time it takes for the particle or the process that begins its journey ($t=0$) from an initial coordinate ($x_0$) to reach either of the specific targets (absorbing boundaries) for the first time \cite{redner_2001}. To estimate ($\langle \tau \rangle$), it is convenient to express the underlying process in terms of a backward FPE, and hence to study the evolution of the survival probability in Laplace space. For a process corresponding to the forward FPE (Eq.~\ref{eqn3}), the backward FPE, where the initial position ($x_0$) is treated as a variable, can be written as \cite{risken_1989,redner_2001,gardiner_1985}:
\begin{equation}
    \frac{\partial Q(t|x_0)}{\partial t}=\left(\theta-x_0\right)\frac{\partial Q(t|x_0)}{\partial x_0}+D_0x_0 \frac{\partial^2 Q(t|x_0)}{\partial x_0^2}.
    \label{eqn10}
\end{equation}
 The survival probability, $Q(t|x_0)$, is the probability that the particle (or the process) starting from an initial position $x_0$ has not reached any of the absorbing boundaries until time $t$. Mathematically, it connects to the probability distribution function as $Q(t|x_0)=\int_{\Omega}p(x,t|x_0)dx$, where $\Omega$ is the allowed diffusion range. In this study, the escape process is confined between two absorbing boundaries at $x_L$ and $x_R$ ($\Omega\in[x_L,x_R]$). Taking the Laplace transformation of both sides, Eq.~\ref{eqn10} can be rearranged into the Kummer type equation (Appendix  \ref{apdA}) that has a general solution \cite{Arfken_1985, Mathews_2022}:
\begin{equation}
    \tilde{Q}(s|x_0)=\frac{1}{s}+A(s)M_0(s) + B(s)U_0(s),
    \label{eqn11}
\end{equation}
where $\tilde{Q}(s|x_0)=\int_{0}^{\infty} e^{-st} Q(t|x_0)dt$, $A$, and $B$ are arbitrary constants that can be evaluated by applying boundary conditions (Appendix \ref{apdA}). Here
\begin{equation}
M_i(s)=M(s,b,z_i),
\end{equation}
and
\begin{equation}
U_i(s)=U(s,b,z_i)
\end{equation}
are the Kummer and Tricomi confluent hypergeometric functions, respectively, with \(b=\frac{\theta}{D_0}\), \(z=\frac{x}{D_0}\), \(z_0=\frac{x_0}{D_0}\), \(z_L=\frac{x_L}{D_0}\), \(z_R=\frac{x_R}{D_0}\), and $i\in\{0,L,R\}$.
The survival probability is connected to the first passage distribution function ($f(t)$) as $\int_0^tf(u)du=1-Q(t|x_0)$. Equivalently, $\tilde{f}(s)=1-s\tilde{Q}(s|x_0)$, which yields
\begin{equation}
    \tilde{f}(s)=-s\left[A(s)M_0(s)+B(s)U_0(s)\right].
    \label{eqn12}
\end{equation}
Now, any moment of first passage time can be derived from $\tilde{f}(s)$ by performing an appropriate differentiation using the Laplace-transform identity for moments: $\langle\tau^n\rangle=(-1)^n\frac{d^n\tilde{f}(s)}{ds^n}|_{s=0}$ \cite{gardiner_1985}. The mean first moment ($n=1$) is connected to $\tilde{f}(s)$ as $\langle\tau\rangle=-\frac{d\tilde{f}(s)}{ds}|_{s=0}$, and the second moment ($n=2$) is related to it as $\langle \tau^2\rangle=\frac{d^2\tilde{f}(s)}{ds^2}|_{s=0}$. Finally, using  Eq.~\ref{eqn12}, one can obtain $\langle\tau\rangle$ as:
\begin{equation}
\langle \tau\rangle 
=\frac{M'_{0}\left(U'_{L}-U'_{R}\right)+M'_{L}\left(U'_{R}-U'_{0}\right)+M'_{R}\left(U'_{0}-U'_{L}\right)}{\left(M'_{L}-M'_{R}\right)-\left(U'_{L}-U'_{R}\right)}.
\label{eqn13}
\end{equation}
 $M'_{i}$ and $ U'_{i}$ denote the first derivatives of the Kummer and Tricomi confluent hypergeometric functions with respect to their first argument, respectively:
\begin{equation}
    M'_{i}=\frac{\partial}{\partial s}M\left(s,b,z_i\right)\big|_{s=0},
\end{equation}
and 
\begin{equation}
    U'_{i}=\frac{\partial}{\partial s}U\left(s,b,z_i\right)\big|_{s=0}.
\end{equation}
\begin{figure}[H]
    \centering
    \includegraphics[width=1\linewidth]{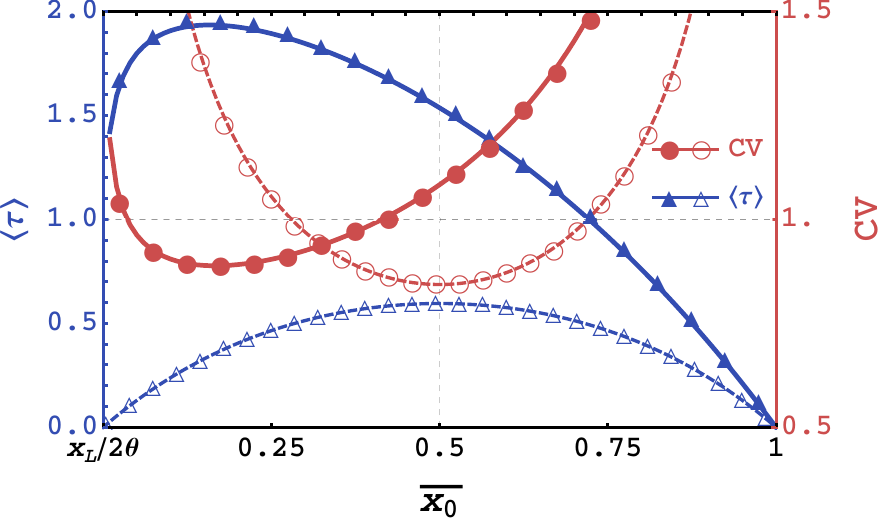}
    \caption{Comparison plot between homogeneous (open markers) and inhomogeneous (filled markers) noise strength showing the dependence of MFPT (left $Y-$axis) and CV (right $Y-$axis)  as a function of scaled initial position $\overline{x_0}=\frac{x_0}{2\theta}$. Results are obtained for $\theta=1$, $D=1$ (constant diffusion term for O-U process), $D_0=1$, with absorbing boundaries located at $x_L=10^{-4}$, and $x_R=2\theta$. Grey colored vertical dashed line is a reference line for $\overline{x_0}=0.5$.}
    \label{fig3}
\end{figure}

\begin{figure}
    \centering
    \includegraphics[width=1\linewidth]{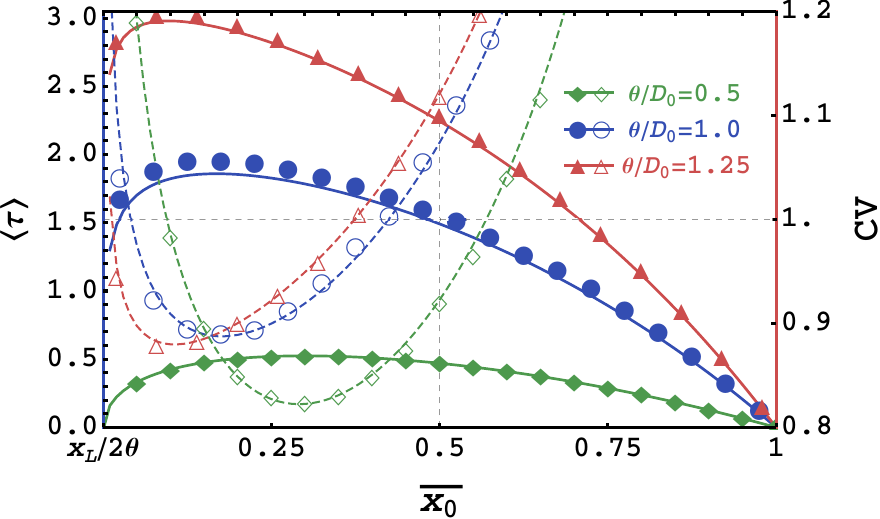}
    \caption{Variation of $\langle\tau\rangle$ (left side of $Y-$axis) and coefficient of variation $(CV)$ (right side of $Y-$ axis) for three different drift-to-noise strength ratios $(\theta/D_0)$ as a function of scaled initial position $\overline{x_0}$. Two fixed absorbing boundaries are taken at $x_L=10^{-4}$ and $x_R=2\theta$, respectively. Solid and dashed lines represent the analytical variation of $\langle\tau\rangle$ and $CV$, respectively. Filled and open markers denote the associated numerical simulated results from the Langevin Eq.~\ref{eqn2}. Grey colored vertical dashed line is a reference line for $\overline{x_0}=0.5$.}
    \label{fig4}  
\end{figure}

 Now, we analyze the impact of inhomogeneity in fluctuation by comparing the mean escape time ($\langle\tau\rangle$) of a Feller process and that of a similar escape event in the presence of homogeneous diffusivity (Ornstein-Uhlenbeck (OU) process) \cite{redner_2001} as a function of the scaled initial position $\overline{x_0} =\frac{x_0}{2\theta}$. The variations are shown in Fig.~\ref{fig3}. The detailed setup and the required expressions for $\langle\tau\rangle$ and $CV$ for an equivalent OU process are represented in Appendix \ref{apdB}. For an OU process, as one expects, a symmetric variation of $\langle\tau\rangle$ as a function of the scaled initial position ($\overline{x_0} $) occurs, with a maximum situated at an equidistance from the two absorbing boundaries. However, inhomogeneous diffusivity breaks this symmetry. The maximum is now shifted towards the frozen-diffusivity side from the potential minima ($\theta$). For the parameter matching used in Fig.~\ref{fig3} (in particular, $D=D_0=1$ and identical absorbing boundaries), we find that $\langle\tau\rangle$ for inhomogeneous diffusivity is larger than that of the corresponding homogeneous-noise (OU) setup at the same initial position. Whenever the particle (or the process) approaches the extinction side, it experiences relatively lower diffusivity, which leads to an increase in the overall $\langle\tau\rangle$ value. Additionally, the frozen diffusivity towards the extinction site leads to less escape through these absorbing boundaries (due to the relatively stronger drift force towards the potential minima). This, in turn, leads to a maximum $\langle\tau\rangle$ at a scaled position inclined towards the extinction site. To support these arguments, we then study the variation of $\langle \tau \rangle$  with scaled initial position ($\overline{x_0}$) for three different regimes of $\frac{\theta}{D_0}$ ratio shown in Fig. \ref{fig4}. As the $\frac{\theta}{D_0}$ ratio increases, the Feller process shows transitions from diffusion-dominated to drift-controlled. For a given noise strength ($D_0$), a higher $\frac{\theta}{D_0}$ indicates a deeper potential well between two absorbing boundaries. Consequently, $\langle \tau \rangle$ at any given $\overline{x_0}$ increases with a higher $\frac{\theta}{D_0}$ ratio to overcome a higher potential threshold to reach the outbreak or extinction point. The left shift of the location of the maxima with higher $\theta$ is consistent with the argument used to explain Fig.~\ref{fig3}.\\


\begin{figure}
    \centering
    \includegraphics[width=0.9\linewidth]{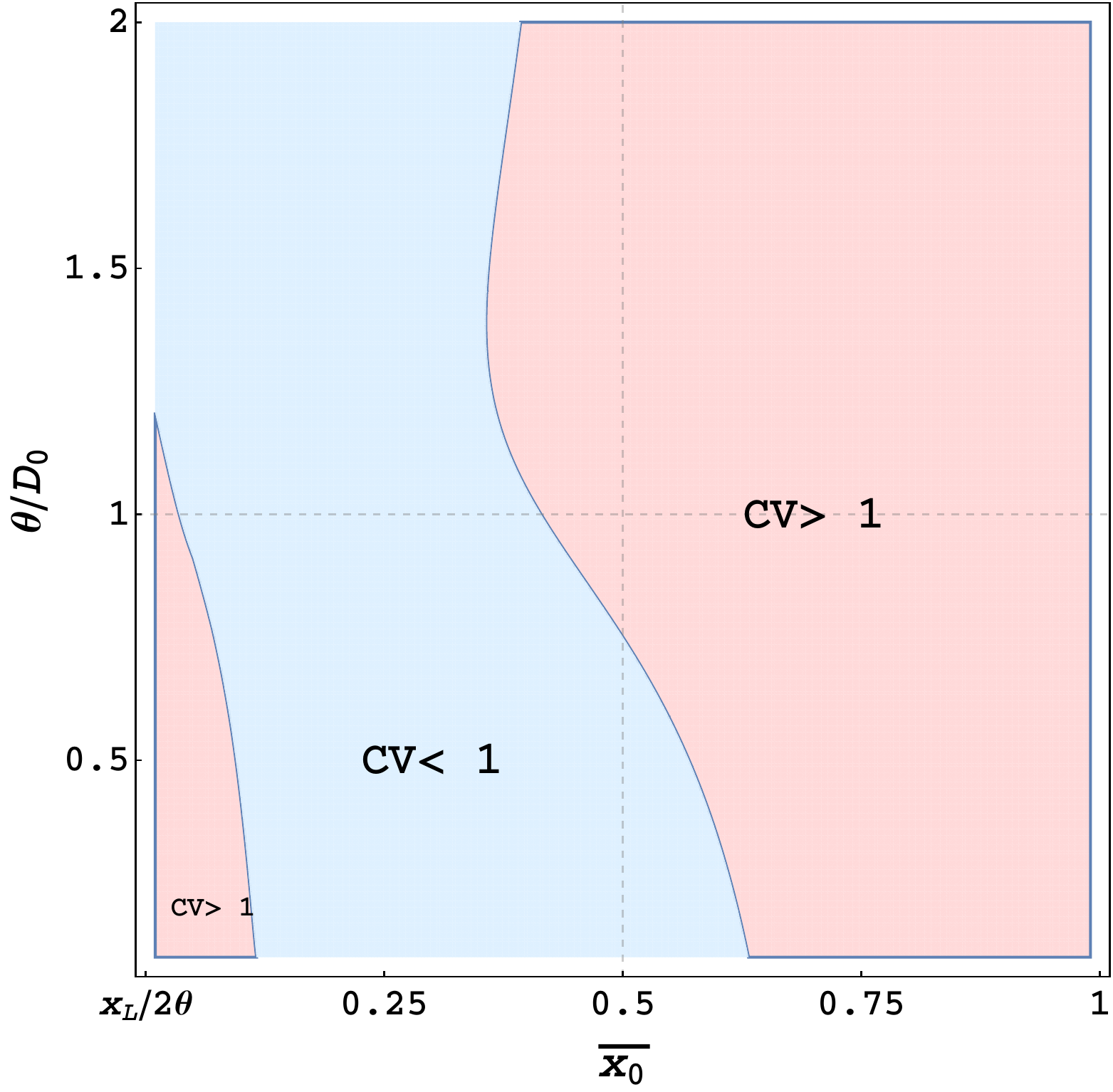}
    \caption{Phase diagram of coefficient of variation $(CV)$ with $\theta/D_0$ and $\overline{x_0}=\frac{x_0}{2\theta}$. Two absorbing boundaries are taken at $x_L=10^{-4}$ and $x_R=2\theta$. The mean value of the first passage time dominates over the fluctuations near the origin for a high $\theta/D_0$ ratio. The dashed lines mark representative reference values of $\overline{x_0}$ and $\theta/D_0$}
    \label{fig5}  
\end{figure}

The coefficient of variation ($CV$) measures the relative fluctuation and is defined as the ratio between the fluctuation and the mean of an observable (here, escape time), $CV=\frac{\sigma(\tau)}{\langle \tau\rangle}$ where $\sigma(\tau)=\sqrt{\langle \tau^2\rangle-\langle \tau\rangle^2}$. $CV$ thus indicates whether fluctuation dominates over the mean of the physical property under consideration. One of the hallmarks of a diffusive (or drift-diffusion) escape event from an interval in the presence of homogeneous fluctuation is that a maximum in $\langle\tau\rangle$ leads to a minimum in $CV$. We find a similar correspondence between the variation of $\langle \tau \rangle$ and $CV$ as a function of the scaled initial position $\overline{x_0}$ in the Feller process, as shown in Figs.~\ref{fig3}-\ref{fig4}. Therefore, unlike homogeneous diffusion, the $CV$ is also found to be least in a position closer to the extinction boundary (not at the potential minima). Figs.~\ref{fig3}-\ref{fig4} also depict that $CV$ is less than unity for a certain range of the initial position. However, fluctuations in escape time dominate its mean ($CV>1$) once the magnitude of the $\frac{\theta}{D_0}$ ratio is appropriately altered. To analyze such dominance, we present a phase diagram (Fig.~\ref{fig5}) as a function of the $\frac{\theta}{D_0}$ ratio and $\overline{x_0}$. For example, for an initial position close to the potential minima ($\overline{x_0} \sim \frac{1}{2}$), the fluctuation is less (dominated by the mean, $CV<1$) with a low drift-to-noise strength ratio (low $\frac{\theta}{D_0}$). However, as the magnitude of $\frac{\theta}{D_0}$ increases, the fluctuations start to dominate. This information is helpful if one is interested in controlling the fluctuation of the escape process by applying a noise control mechanism like stochastic resetting \cite{satya_2011PRL,shlomi_2016PRL,somrita_2019JPA,gregory_2020JPAMT,somrita_2020JCP,somrita_2023jcp,somrita_2025PRL,DMondal_2022jpamt}.

\section{Splitting probability: Extinction or Outbreak}
\label{sec5}
\begin{figure}
    \centering
    \includegraphics[width=1\linewidth]{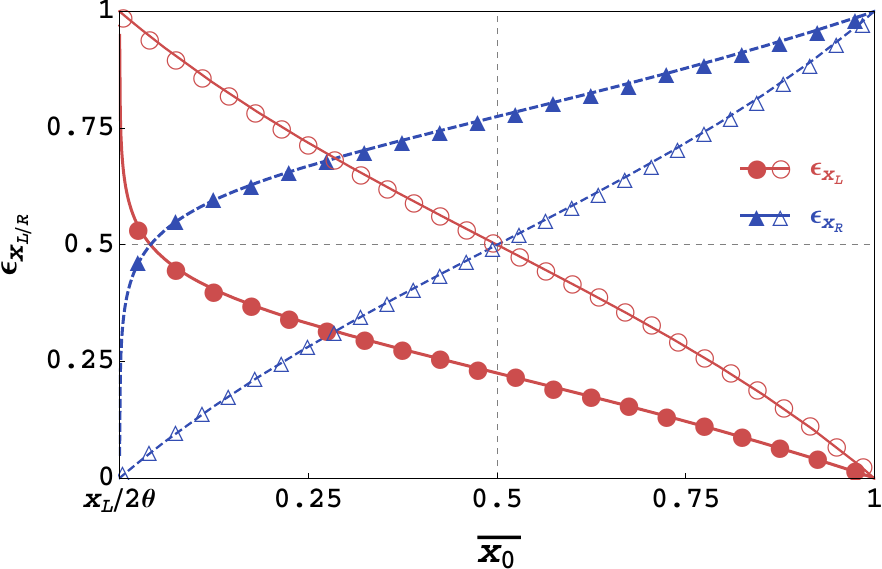}
    \caption{Comparison plot between homogeneous (open markers) and inhomogeneous (filled markers) noise strength showing the dependence of splitting probabilities ($\epsilon_{x_L}$ or $\epsilon_{x_R}$)  as a function of scaled initial position $\overline{x_0}=\frac{x_0}{2\theta}$. Results are obtained for $\theta=1$, $D=1$, $D_0=1$, with absorbing boundaries located at $x_L=10^{-4}$, and $x_R=2\theta$. The gray colored dashed lines represent reference values of $\overline{x_0}$ and $\epsilon_{x_{L/R}}$.}
    \label{fig6}
\end{figure}

For an escape problem from an interval, it is always important to understand which boundary the particle is more likely to absorb into and how this absorption probability depends on the system parameter. In this context, a useful variable, namely the splitting probability, is defined as the likelihood of the stochastic process absorbing to one particular target boundary (say $x_L$) before reaching the other available boundaries (here $x_R$) or vice versa \cite{redner_2001,gardiner_1985}. For a stochastic escape under uniform fluctuations, starting from an initial position equidistant to either absorbing boundary, the splitting probabilities to both boundaries are equal (as shown in Fig.~\ref{fig6}) for obvious reasons. Also, the change in splitting probabilities to either side is symmetric if a spatial initial bias is introduced. Now, the pertinent question is, in the presence of spatially heterogeneous fluctuations, where does the particle absorb more? In this spirit, we derive its splitting probability using the following relation \cite{gardiner_1985}:
\begin{equation}
    \epsilon_{x_L}=\frac{\int_{x_0}^{x_R}\phi(x)dx}{\int_{x_L}^{x_R}\phi(x)dx},
    \label{eqn11a}
\end{equation}
and $\epsilon_{x_R}+\epsilon_{x_L} =1$. Here $\epsilon_{x_L}$ and $\epsilon_{x_R}$ denote the splitting probabilities towards the left (extinction probability) and right (outbreak probability) absorbing boundaries, respectively. In the present context, the potential function: $\phi(x)=e^{\frac{x}{D_0}}x^{-\frac{\theta}{D_0}}$. We solve Eq.~\ref{eqn11a} to obtain the following exact theoretical prediction in terms of incomplete gamma functions:
\begin{equation}
    \epsilon_{x_L}=\frac{\Gamma\left(1-\frac{\theta}{D_0},-\frac{x_R}{D_0}\right)-\Gamma\left(1-\frac{\theta}{D_0},-\frac{x_0}{D_0}\right)}{\Gamma\left(1-\frac{\theta}{D_0},-\frac{x_R}{D_0}\right)-\Gamma\left(1-\frac{\theta}{D_0},-\frac{x_L}{D_0}\right)}.
    \label{eqn17}
\end{equation}
Here, the upper incomplete gamma function is defined as $\Gamma(a,x)=\int_x^\infty t^{a-1} e^ {- t} dt$.\\

\begin{figure}
    \centering
    \includegraphics[width=0.9\linewidth]{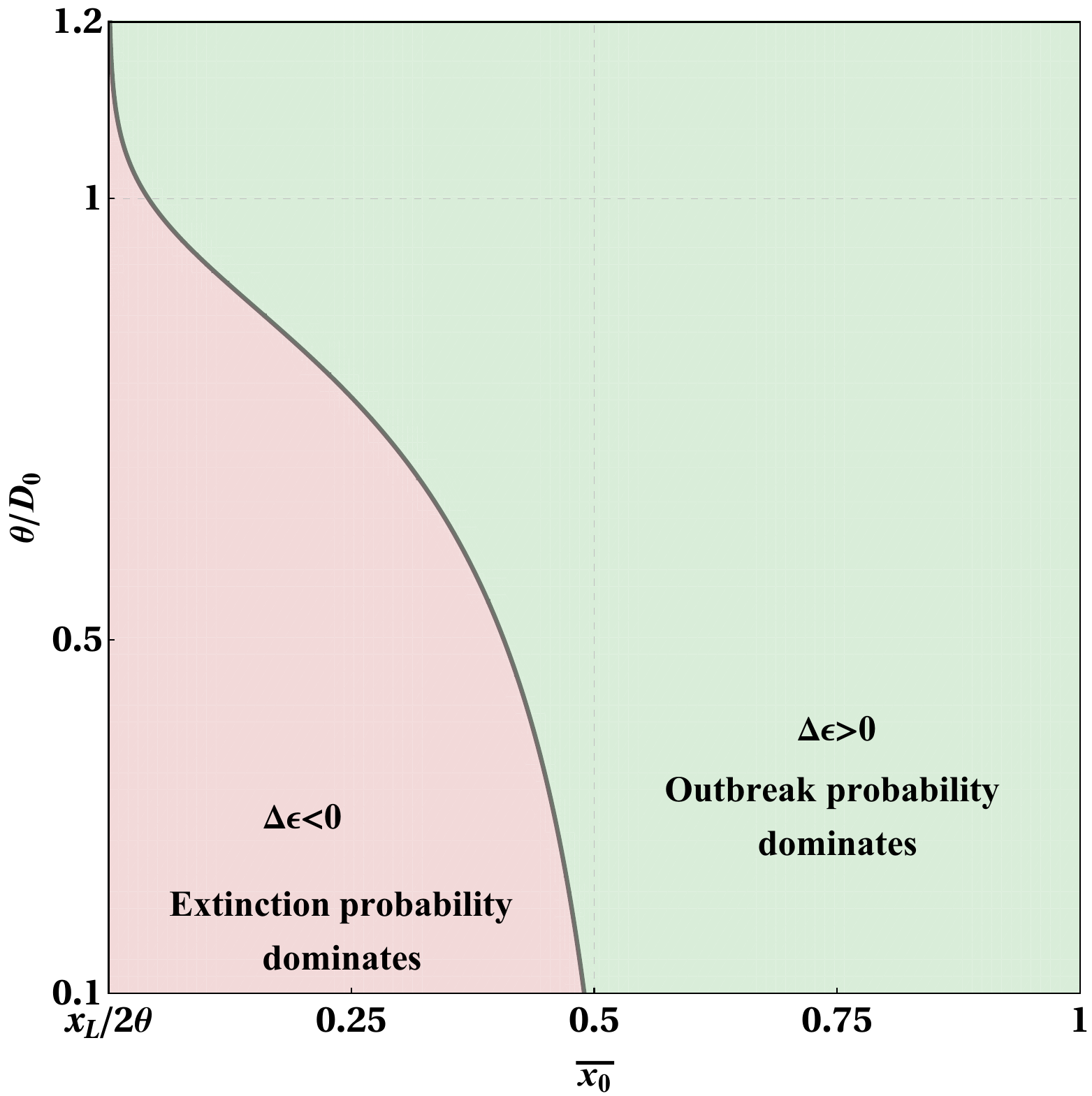}
    \caption{Phase diagram of the splitting probability difference \(\Delta\epsilon=\epsilon_{x_R}-\epsilon_{x_L}\) for different values of $\frac{\theta}{D_0}$ and scaled initial position \(\overline{x_0}\). Two reference absorbing boundaries are positioned at $x_L=10^{-4}$ and $x_R=2\theta$. The light gray dashed lines mark reference values of $\overline{x_0}$ and $\frac{\theta}{D_0}$.}
    \label{fig7}
\end{figure}

In Fig.~\ref{fig6}, we explore the effect of spatially heterogeneous noise on the extinction and outbreak probabilities as a function of the scaled initial position $\overline{x_0}$. The variation shows that inhomogeneity in fluctuations leads to a lower extinction probability (a higher outbreak possibility) than in an analogous O-U setup for any given initial position. Because of the almost frozen diffusivity near the extinction boundary, the particles are pushed by drift towards the potential minimum. On the other hand, the outbreak threshold is more likely to be reached as the stochastic force becomes increasingly strong near its threshold. Therefore, although a relatively frozen fluctuation zone slows the escape process, a higher fraction of particles is absorbed at the outbreak site. Consequently, an equal chance of extinction and outbreak possibility arises when the process starts its initial journey from a position closer to the extinction site. To better understand when the process is driven by outbreak (or extinction) phenomena and how this dominance depends on system parameter conditions, we have also shown a phase diagram of the splitting probability difference \(\Delta\epsilon=\epsilon_{x_R}-\epsilon_{x_L}\) in Fig.~\ref{fig7}. The variation shows that, irrespective of \(\frac{\theta}{D_0}\) values, the process always favors outbreak outcomes if it is initialized from an unbiased position \(\overline{x_0}=\frac{1}{2}\). Also, extinction phenomena are dominated only if we start closer to that boundary, as well as if the drift-to-noise strength ratio is low (roughly $\frac{\theta}{D_0}\le1$). The result is consistent to the stationary PDF as shown in Fig.~\ref{fig2}.\\


\begin{figure}
    \centering
    \includegraphics[width=0.9\linewidth]{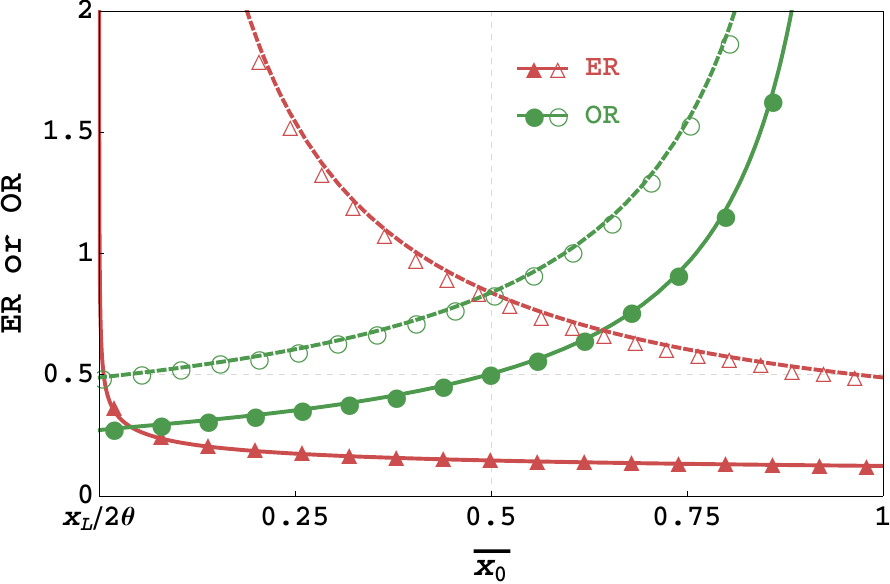}
    \caption{Effective outbreak (circular points) or extinction (triangular points) rates comparison between O-U (open markers) and Feller diffusion (filled markers) process. The dashed lines mark representative reference values of $\overline{x_0}$ and effective rates.}
    \label{fig8}
\end{figure}

So far, we have explored how mean escape time ($\langle\tau\rangle$) and splitting probabilities ($\epsilon_{x_L/x_R}$) in a bounded Feller process are influenced by the drift-to-noise strength ratio in a nontrivial manner. The inhomogeneous noise structure often favors an outbreak over extinction but slows the overall escape process. While $\langle\tau\rangle$ defines the characteristic mean lifetime of the process starting from $x_0$, $\epsilon_{x_L/x_R}$ indicates the likelihood of a target-based outcome for the same. Therefore, a ratio between $\epsilon_{x_L/x_R}$ and $\langle\tau\rangle$ may provide an effective outcome-specific rate. Accordingly, we define the effective extinction rate $\left(ER= \frac{\epsilon_{x_L}}{\langle\tau\rangle}\right)$ and the outbreak rate $\left(OR=\frac{\epsilon_{x_R}}{\langle\tau\rangle}\right)$. These rates are analogous to reward rates in decision-making theory or speed-accuracy trade-off ratios in stochastic kinetics (e.g., enzymatic kinetics) with multiple outcomes \cite{jonathan_2006pr,shadlen_2007arn,leibler_2012pnas,peliti_2015jsm,igoshin_2017pnas}.  In Fig.~\ref{fig8}, we depict the effect of the inhomogeneous fluctuation on the effective outbreak rate $OR$ (Extinction rate $ER$) as a function of the scaled initial position. A lift in the symmetric variation is observed. The effective outbreak rate now dominates the extinction rate, even if the process starts from an initial position close to the extinction site. Therefore, the inhomogeneous fluctuation leads to a higher effective outbreak rate than extinction in a decision-making scenario. However, the effective outbreak (or extinction) rate is lower than that of an analogous O-U process due to the slower escape rate. 
\section{Conclusions}
\label{sec7}
We present a comprehensive study of the mean first-passage time for a Feller process in an interval using an exactly solvable model. We map the exit windows of the interval as extinction and outbreak sites, respectively, concerning the position coordinates with the same potential energy. We set the extinction boundary close to zero, where the noise fluctuations freeze. The outbreak site experiences stronger fluctuations and is located at a position twice that of the stablest energy point. \\

A spatially inhomogeneous fluctuation leads to asymmetric escape events. Consequently, we find a nontrivial variation of the mean exit time and the splitting probability as a function of initial position and the strength of the drift-to-noise ratio. The inhomogeneous fluctuations increase the outbreak probability, even if the process starts from an unbiased initial state, due to a frozen bottleneck near the extinction point. We also observe that the mean escape time reaches a maximum when the Feller particle is initialized in a position inclined to the hell (extinction) site. The ratio of splitting probabilities and the mean escape time show a better effective outbreak rate than extinction.\\
 
Finally, the condition of maximum exit time is then nicely corroborated with the least $CV$, which is the hallmark for stochastic escape from an interval. Depending on the relative values of the drift-to-noise strength and the initial bias, fluctuations in exit time may dominate its mean ($CV$ greater than unity) or vice versa. Therefore,  the stochastic escape from confinement in a Feller diffusion process can be further extended to explore the effect of stochastic resetting. Such an extension will enhance the applicability of this model to a broader class of physical and biological systems and will be addressed elsewhere.\\

\section*{Acknowledgments}
P.B. acknowledges IIT Tirupati and DST/INSPIRE for the fellowship (Project No. DST/INSPIRE/03/2023/002266). D.M. thanks SERB (Project No. ECR/2018/002830/CS), Department of Science and Technology, Government of India, for financial support.
\section*{Data Availability}
The data supporting all the findings of this study are available in this article or in the appendices.
\appendix
\section{Derivation of MFPT}
\label{apdA}
Performing a Laplace transformation on both sides of Eq.~\ref{eqn10}, one can reach the following equation:
\begin{equation}
    D_0x_0\frac{d^2\tilde{Q}(s|x_0)}{d x_0^2}-(x_0-\theta)\frac{d \tilde{Q}(s|x_0)}{d x_0}-s\tilde{Q}(s|x_0)=-1.
    \label{eqna3}
\end{equation}
Clearly, the above ODE is inhomogeneous. Converting this to a homogeneous ODE with a suitable new variable $\tilde{q}(s|x_0)=\tilde{Q}(s|x_0)-\frac{1}{s}$, we will obtain
\begin{equation}
    D_0x_0\frac{d^2\tilde{q}(s|x_0)}{d x_0^2}+(\theta-x_0)\frac{d \tilde{q}(s|x_0)}{d x_0}-s\tilde{q}(s|x_0)=0.
    \label{eqna2}
\end{equation}
Introducing a dimensionless variable $z_0=\frac{x_0}{D_0}$ and using the chain rule \(\frac{d}{dx_0}=\frac{1}{D_0}\frac{d}{dz_0}\), one can find the following derivative relations:
\begin{equation}
    \frac{d\tilde{q}(s|x_0)}{dx_0}=\frac{1}{D_0}\frac{d\tilde{q}(s|x_0)}{dz_0};
\end{equation}
\begin{equation}
    \frac{d^2\tilde{q}(s|x_0)}{dx_0^2}=\frac{1}{D_0^2}\frac{d^2\tilde{q}(s|x_0)}{dz_0^2}.
    \label{eqn30}
\end{equation}
Putting these forms back in Eq.~\ref{eqna2}, we rearrange it as follows: 
\begin{equation}
    z_0\frac{d^2\tilde{q}(s|z_0)}{d z_0^2}+\left(\frac{\theta}{D_0}-z_0\right)\frac{d \tilde{q}(s|z_0)}{dz_0}-s\tilde{q}(s|z_0)=0.
\end{equation}
The equation above represents Kummer's equation. The general solution of the equation (in Laplace space) takes the form:
\begin{equation}
    \tilde{q}(s|z_0)=AM\left(s,b, z_0\right)+BU\left(s,b, z_0\right),
\end{equation}
where $M(s,b,z_0)$ and $U(s,b,z_0)$ are the confluent hypergeometric functions of the first and second kinds, respectively, with the following definitions:
\begin{equation}
    M\left(s,b,z_0\right)=\sum_{n=0}^{\infty}\frac{(s)_n}{(b)_n}\frac{z_0^n}{n!}
\end{equation}
and
\begin{equation}
\begin{split}
    U\left(s, b, z_0\right) ={} & \frac{\Gamma(1-b)}{\Gamma(1+s-b)} M\left(s, b, z_0\right) \\
    & + \frac{\Gamma(b-1)}{\Gamma(s)} z_0^{1-b} M\left(s, b, z_0\right).
\end{split}
\end{equation}
Importantly, $A$ and $B$ are the arbitrary constants that can be evaluated by applying absorbing boundary conditions: $Q(t|x_L)=\tilde{Q}(s|x_L)=0$ and $Q(t|x_R)=\tilde{Q}(s|x_R)=0$, and they take the form:
\begin{eqnarray}
    A&=&\frac{U_{L}(s)-U_{R}(s)}{s\left[U_{R}M_{L}-U_{L}M_{R}\right]},   \nonumber \\
     B&=&\frac{M_{R}(s)-M_{L}(s)}{s\left[U_{R}M_{L}-U_{L}M_{R}\right]},
    \label{eqn36}
\end{eqnarray}
where $M_{i}=M\left(s,b,z_i\right)$ and \(U_{i}=U\left(s,b,z_i\right)\). Now, writing the general solution in terms of the original variable \(\tilde{Q}(s|x_0)\), we can obtain Eq.~\ref{eqn11} of the main text. Hence, using the definition of mean first passage time, one can obtain the closed form of \(\langle\tau\rangle\) as given in Eq.~\ref{eqn13}.\\\\
\section{Derivation related to O-U process}
\label{apdB}
 To contrast the behavior of the Feller process with a system exhibiting homogeneous fluctuations, we consider an Ornstein-Uhlenbeck process (O-U) confined to the same interval. Therefore, an O-U process can be described by the following dimensionless Langevin equation\cite{gardiner_1985} in It\^{o} sense
 \begin{equation}
     \frac{dx}{dt}=(\theta-x)+\sqrt{2D}\eta(t),
 \end{equation}
where \(\theta\) denotes the location of the potential's center. \(\eta(t)\) represents normalized Gaussian white noise with zero mean and delta correlation in time. The strength of the homogeneous fluctuations is given by \(D\), which is often denoted as the diffusion coefficient. To facilitate comparison with the Feller-process setup, we impose two fixed absorbing boundaries as \(x_L\sim0\) and \(x_R=2\theta\).
The first-passage time and escape probabilities through a prescribed region for this configuration are well established and extensively discussed in the literature \cite{redner_2001}. The mean escape time for a particle under an OU process within a finite interval (starting from an initial position ($x_0$)) can be obtained in exact, compact, and standard integral form \cite{redner_2001} as:
\begin{equation}
    \langle\tau\rangle =\frac{ \mathcal{Z}(x_0)-\mathcal{Z}_2(x_0) }{D\int_{x_L}^{x_R}\frac{dy}{\psi(y)}}
\end{equation}
where $\mathcal{Z}(x_0)=\Big(\int_{x_L}^{x_0}\frac{dy}{\psi(y)}\Big)\int_{x_0}^{x_R}\frac{dy'}{\psi(y')}\int_{x_L}^{y'}\psi(z)dz$ and $\mathcal{Z}_2(x_0)=\Big(\int_{x_0}^{x_R}\frac{dy}{\psi(y)}\Big)\int_{x_L}^{x_0}\frac{dy'}{\psi(y')}\int_{x_L}^{y'}\psi(z)dz$ with $\psi(x)=\exp{\big[-\frac{1}{D}(\frac{x^2}{2}-\theta x)\big]}$. 
One can also consider the Backward Fokker-Planck approach, and following a similar method as described in Appendix \ref{apdA}, we obtain the mean escape time in closed form.
\begin{equation}
\langle \tau\rangle 
=\frac{H'_{R}\left(M'_{0}-M'_{L}\right)+H'_{L}\left(M'_{R}-M'_{0}\right)+H'_{0}\left(M'_{L}-M'_{R}\right)}{2\left(H'_{L}-H'_{R}\right)+M'_{L}-M'_{R}},
\end{equation}
where $ M'_{i}$ denotes the first derivatives of the confluent hypergeometric functions with respect to their first argument, and $ H'_{i}$ is a Hermite function. The derivatives are expressed as follows
\begin{align}
    M'_{i}=&\frac{\partial}{\partial s}M\bigg(s,\frac{1}{2},\frac{\left(x_i-\theta\right)^2}{2D}\bigg)\bigg|_{s=0},   \nonumber \\
    H'_{i}=&\frac{\partial}{\partial s}H\bigg(s,\frac{x_i-\theta}{\sqrt{2D}}\bigg)\bigg|_{s=0},
\end{align}
with $i\in\{0,L,R\}$. Following a similar approach (survival probability in Laplace space), one can also obtain a closed form of the $CV$ by calculating the second moment of the first passage time (not shown here as the expression is lengthy and not very aesthetically pleasing!). Finally, denoting $\epsilon_{x_L}$ and $\epsilon_{x_R}$ as the splitting probabilities towards the left (extinction) and right (outbreak) absorbing boundaries, and following the definition Eq.~\ref{eqn11a}, we obtain:
\begin{equation}
    \epsilon_{x_L}=\frac{erfi\left(\frac{x_R-\theta}{\sqrt{2D}}\right)-erfi\left(\frac{x_0-\theta}{\sqrt{2D}}\right)}{erfi\left(\frac{x_R-\theta}{\sqrt{2D}}\right)-erfi\left(\frac{x_L-\theta}{\sqrt{2D}}\right)},
\end{equation}
and $\epsilon_{x_R}=1-\epsilon_{x_L}$. Here $erfi(z)$ gives the imaginary error function.

\bibliographystyle{apsrev4-2}
\bibliography{refs}

\end{document}